\documentclass[twocolumn,showpacs,preprintnumbers,amsmath,amssymb]{revtex4}
\usepackage{graphicx}
\usepackage{dcolumn}
\usepackage{bm}

\begin{document}

\title{Bogoliubov Excitations in a Kronig-Penney Potential}

\author{Ippei Danshita}
\email{danshita@kh.phys.waseda.ac.jp}
\author{Susumu Kurihara}
\affiliation{Department of Physics, Waseda University, \=Okubo, Shinjuku, Tokyo 169-8555, Japan}
\author{Shunji Tsuchiya}
\email{stuchiya@iis.u-tokyo.ac.jp}
\affiliation{Institute of Industrial Science, University of Tokyo, Komaba, Meguro, Tokyo 153-8505, Japan}
\date{\today}

\begin{abstract}
With use of the Kronig-Penney model, we study the excitation spectrum of a Bose-Einstein condensate in a one-dimensional periodic potential.
We solve the Bogoliubov equations analytically and obtain the band structure of the excitation spectrum for arbitrary values of the lattice depth.
We find that the excitation spectrum is gapless and linear at low energies, and that it is due to the {\it anomalous tunneling} of low energy excitations, predicted by Kagan {\it et al.}.
\end{abstract}

\maketitle

\section{\label{sec:level1}Introduction}

In studies of Bose-Einstein condensation in ultracold atomic gases, elementary excitations play an important role in understanding various properties of the Bose-Einstein condensates, such as dynamics, thermodynamics and superfluidity.
Elementary excitations have been experimentally investigated in trapped Bose-Einstein condensates, including the observation of collective modes~\cite{rf:BEC,rf:jila,rf:mit} and the measurement of the Bogoliubov excitation spectrum with use of Bragg spectroscopy~\cite{rf:BEC,rf:bogo}.

An optical lattice is a periodic potential for atoms created by standing waves of laser beams.
Experimental observations have also revealed properties of elementary excitations of Bose-Einstein condensates in an optical lattice.
St\"oferle {\it et al.} have studied excitation spectra of Bose-Einstein condensates in an optical lattice~\cite{rf:sf_mi}.
The Bose-Einstein condensates were excited by a modulation of the lattice depth, and a broad continuum of the excitation spectrum was observed; they suggested that the broad continuum characterizes the superfluid phase.

In the present paper, we study elementary excitations of Bose-Einstein condensates in an optical lattice with use of a Kronig-Penney potential.
In previous theoretical papers~\cite{rf:origin,rf:dynam,rf:tight,rf:DNLS,rf:comp,rf:finite,rf:sounv,rf:phd,rf:soft}, elementary excitations in a sinusoidal lattice potential have been studied.
Berg-S\o rensen {\itshape et} {\itshape al}. numerically solved the Bogoliubov equations for a one-dimensional optical lattice, and they also calculated the low energy excitations analytically for shallow lattices within the Thomas-Fermi limit; they showed that the excitation spectrum is phonon-like at low energies~\cite{rf:origin}.
The phonon dispersion of the excitation spectrum is directly connected to the superfluidity of a Bose-Einstein condensate in an optical lattice~\cite{rf:land}, and most of the papers support the phonon dispersion.
In contrast, Ichioka {\itshape et} {\itshape al}. have reported on softening of the low energy excitations for deep lattices by numerically solving the Bogoliubov equations~\cite{rf:soft}.
Such softening of the excitation spectrum, if it really exists, suggests an instability of superfluidity of a Bose-Einstein condensate in an optical lattice. 
One of our purposes is to investigate the possibility of the softening by analytically solving the Bogoliubov equations with a Kronig-Penney potential.
We will show that the softening does not occur in the Kronig-Penney model.

In previous papers, analytical methods to calculate the excitation spectrum in the presence of a sinusoidal lattice potential are limited either to deep or shallow lattices.
For deep lattices where the overlap of the condensate wave functions in neighboring sites is sufficiently small, a tight-binding approximation provides analytical expressions for the first band of the excitation spectrum~\cite{rf:tight,rf:DNLS,rf:comp,rf:finite}.
For shallow lattices, the perturbative treatment of the lattice potential is applicable~\cite{rf:sounv,rf:phd}.
An advantage of the Kronig-Penney model is that one can calculate the excitation spectrum exactly with analytic methods for arbitrary values of the lattice depth.

Another advantage of the Kronig-Penney model is that the excitation spectrum can be related to tunneling properties of the excitations through a single potential barrier.
Kagan {\itshape et} {\itshape al}. studied the tunneling problem of excitations and predicted that a barrier is transparent for excitations within a limited range of low energies; they called such a behavior the {\it anomalous tunneling}~\cite{rf:antun}.
We shall see, in fact, that the anomalous tunneling is crucial to the phonon-like form of the low energy excitations and that the softening of the excitation spectrum does not occur.

The outline of the present paper is as follows.
In Sec. \!II, we introduce a formulation of the problem using the Bogoliubov theory and calculate the condensate wave function in a Kronig-Penney potential.
In Sec. \!III, we analytically solve the Bogoliubov equations and obtain the band structure of the excitation spectrum in the Kronig-Penney potential.
We derive the phonon-like form of the low energy excitation.
We also calculate the phonon velocity and the band gap.
Conclusions and open questions are discussed in the final section.

\section{Condensate wave function in a Kronig-Penney potential}
We consider a Bose-Einstein condensate confined in a combined potential of an axisymmetric harmonic potential and a one-dimensional periodic potential along the axial direction (the $x$ axis).
As the periodic potential, we adopt a Kronig-Penney potential,
  \begin{eqnarray}
  V(x)=V_0\sum_{n=-\infty}^{\infty}\delta(x-na),
  \end{eqnarray}
where $a$ is the lattice constant, and $V_0$ is the strength of the $\delta$-function potential barrier, expressing the lattice depth.
It is assumed that the condensate is so elongated along the axial direction that the axial confinement can be neglected.
We assume that the frequency of the harmonic potential of the radial direction is large enough compared to the excitation energy for the axial direction.
In this situation, the one-dimensional treatment of the problem is justified.

Our formulation of the problem is based on the mean-field theory~\cite{rf:BEC}, which consists of the time-independent Gross-Pitaevskii equation and the Bogoliubov equations.
They are
\begin{eqnarray}
    \Bigl[-\frac{\hbar^2}{2m}\frac{d^2}{dx^2}+V(x)+g|\Psi_0(x)|^2\Bigr]
    \Psi_0(x) = \mu\Psi_0(x),
    \label{eq:sGPE}
  \end{eqnarray}
and
       \begin{eqnarray}
           \begin{array}{cc}
               \left(
                 \begin{array}{cc}
                 H_0 & -g{\Psi_0(x)}^2 \\
                 g{\Psi_0(x)}^{\ast2} & -H_0
                 \end{array}
               \right) 
               \left(
                 \begin{array}{cc}
                 u(x) \\ v(x)
                 \end{array}
               \right)
               = \varepsilon\left(
                 \begin{array}{cc}
                   u(x) \\ v(x)
                 \end{array}
               \right),               
           \end{array}\\ \label{eq:BdGE}
           H_0 = -\frac{\hbar^2}{2m}\frac{d^2}{dx^2}
               -\mu+V\left(x\right)+2g|\Psi_0(x)|^2.
       \end{eqnarray}
Here $\mu$ is the chemical potential and $m$ is the mass of an atom.
Since the radial confinement is harmonic and sufficiently tight, the coupling constant is affected by the harmonic oscillator length $a_{\perp}$ of the radial confinement as $g=\frac{2\hbar^2a_s}{ma_{\perp}^2}$, where $a_s$ is the $s$-wave scattering length~\cite{rf:1dime}.
The Gross-Pitaevskii equation determines the condensate wave function $\Psi_0(x)$.
The Bogoliubov equations determine the energy $\varepsilon$ of the elementary excitations of the condensate and the wave functions of the excitation $\left(u(x), v(x)\right)^{\bf t}$.
The elementary excitations correspond to small fluctuations of the condensate wave function from the static equilibrium~\cite{rf:BEC}.

\begin{figure}[b]
\includegraphics[width=3 in, height=1.2 in]{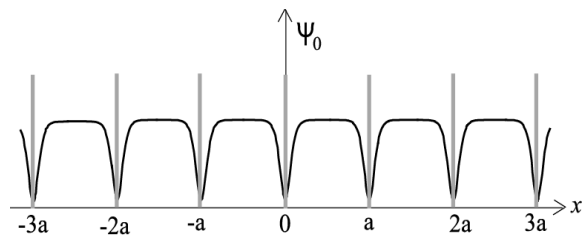}
\caption{\label{fig:kpcon}
The schematic picture of a condensate in a Kronig-Penney potential.
                              }
\end{figure}

The periodic potential is sinusoidal in experiments of atomic gases trapped in an optical lattice.
However, the Kronig-Penney model is useful to understand the problem of the periodic potential qualitatively, because it allows an analytical treatment of the problem.
Schematic picture of the condensate in the Kronig-Penney potential is shown in Fig. \!\ref{fig:kpcon}.


The chemical potential is related to the number of condensate atoms $N_0$ in each well by the normalization condition:
  \begin{equation}
     \int_{na}^{(n+1)a}dx|\Psi_0(x)|^2=N_0.\label{eq:norm}
  \end{equation}

At first, we shall solve Eq. \!(\ref{eq:sGPE}) and obtain the condensate wave function in a Kronig-Penney potential.
We assume that the condensate does not have supercurrent; therefore one can regard, without loss of generality, the condensate wave function as real.
Multiplying Eq. \!(\ref{eq:sGPE}) by $\frac{d\Psi_0}{dx}$, one obtains the first integral of Eq. \!(\ref{eq:sGPE}), 
  \begin{eqnarray}
    \left(\frac{d\Psi_0}{dx}\right)^2 = 
    \frac{\mu}{\xi^2 g}(A^2-\frac{g}{\mu}\Psi_0^2)^2,\label{eq:integ}
  \end{eqnarray}
where $\xi\equiv\frac{\hbar}{\sqrt{m\mu}}$ is the healing length, and $A$ is a constant corresponding to the value of the condensate wave function at the center of each lattice site.
It is expressed as
  \begin{eqnarray}
  A\equiv\sqrt{\frac{g}{\mu}}\Psi_0((n+\frac{1}{2})a).
  \end{eqnarray}
Integrating Eq. \!(\ref{eq:integ}) again, one obtains
  \begin{eqnarray}
    \Psi_0(x)\!=\!\sqrt{\frac{\mu}{g}}A\,
    \mathrm{sn}\!\left(\!\!\sqrt{2-A^2}(|x\!-\!na|\!+\!x_0),\!
    \frac{A}{\sqrt{2-A^2}}\!\right)\! ,\nonumber
  \end{eqnarray}
  \begin{eqnarray}
    \left(n-\frac{1}{2}\right)a<x<\left(n+\frac{1}{2}\right)a,\label{eq:jacobi}
  \end{eqnarray}
The boundary conditions at $x=na$ and $x=(n+\frac{1}{2})a$ determine the constants $A$ and $x_0$.
This solution has been obtained in Refs.~\cite{rf:kron,rf:penn}.

In order to calculate the excitation spectrum analytically, we assume that the lattice constant is sufficiently larger than the healing length.
This assumption also allows us to relate the band structure of the excitation spectrum to the tunneling properties of the excitations.
In this situation, the condensate wave function near the center of each lattice site is not affected by other potential barriers.
Then, one approximately obtains the ground state solution of Eq. \!(\ref{eq:sGPE}) as
  \begin{eqnarray}
     \Psi_0(x) = \sqrt{\frac{\mu}{g}}
                 \mathrm{tanh}\left(\frac{|x-na|+x_0}{\xi}\right),
     \nonumber\\
            \left(n-\frac{1}{2}\right)a<x<\left(n+\frac{1}{2}\right)a,
            \label{eq:cnds}
  \end{eqnarray}
where $x_0$ is determined by the boundary condition at $x=na$
  \begin{eqnarray}
  \Psi_0(na+0)=\Psi_0(na-0), 
  \end{eqnarray}
  \begin{eqnarray}
  \frac{d\Psi_0}{dx}\biggl.\biggr|_{na+0}
  &=&
  \frac{d\Psi_0}{dx}\biggl.\biggr|_{na-0}+\frac{2mV_0}{\hbar^2}\Psi_0(na),
  \end{eqnarray}
as
  \begin{eqnarray}
  {\rm tanh}\frac{x_0}{\xi}=\frac{-V_0+\sqrt{V_0^2+4(\mu\xi)^2}}{2\mu\xi}.
  \end{eqnarray}
This condensate wave function corresponds to the $A\to1$ limit of the Eq. \!(\ref{eq:jacobi}).
\begin{figure}[t]
\includegraphics[width=3 in, height=2 in]{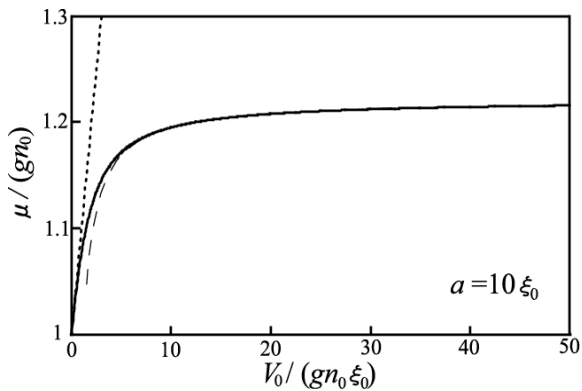}
\caption{\label{fig:chem_pote}
The solid line represents the chemical potential $\mu$ for $a=10\xi_0$, as a function of $V_0$.
The dashed (dotted) line represents an approximate value of $\mu$ when $V_0 \gg gn_0\xi_0$ ($V_0 \ll gn_0\xi_0$).
                              }
\end{figure}

Substituting Eq. \!(\ref{eq:cnds}) into the normalization condition of Eq. \!(\ref{eq:norm}), we derive the relation between the chemical potential and $N_0$:
  \begin{equation}
    \mu\left(1-2\frac{\xi}{a}+2\frac{\xi}{a}
    \mathrm{tanh}\frac{x_0}{\xi}\right)=gn_0,
    \label{eq:chemdet}
  \end{equation}
where $n_0\equiv \frac{N_0}{a}$ is the averaged density of the condensate.
One can obtain approximate solutions of Eq. \!(\ref{eq:chemdet}) in the limits of $V_0 \gg gn_0\xi_0$ and $V_0 \ll gn_0\xi_0$, where $\xi_0 \equiv \frac{\hbar}{\sqrt{mgn_0}}$.
When $V_0 \gg gn_0\xi_0$, we expand Eq. \!(\ref{eq:chemdet}) into power series of $\frac{\xi_0}{a}$ and $\frac{gn_0\xi_0}{V_0}$, and obtain
  \begin{eqnarray}
    \mu \!\simeq\!
    gn_0\!\left(\!1+\frac{2\xi_0}{a}+\frac{2\xi_0^2}{a^2}
    -\frac{2gn_0\xi_0^2}{aV_0}+\frac{\xi_0^3}{a^3}-\frac{6gn_0\xi_0^3}{a^2V_0}
    \!\right)\!. \label{eq:cheml}
  \end{eqnarray}
In a similar way, when $V_0 \ll gn_0\xi_0$, we expand Eq. \!(\ref{eq:chemdet}) into power series of $\frac{\xi_0}{a}$ and $\frac{V_0}{gn_0\xi_0}$, and obtain
  \begin{eqnarray}
    \mu
    \simeq
     gn_0\left(1+\frac{V_0}{gn_0 a}-\frac{V_0^2}{4a\xi_0(gn_0)^2}\right).
    \label{eq:chems}
  \end{eqnarray}
In Eqs. \!(\ref{eq:cheml}) and (\ref{eq:chems}), we express the expansions up to the third order of the small parameters.
We show the chemical potential as a function of the potential strength $V_0$ in Fig. \!\ref{fig:chem_pote}.
The chemical potential increases monotonically as the potential strength increases, because the presence of the potential barriers makes the effect of repulsive interaction more significant.
%
%
\section{Excitation spectrum}
In this section, we shall solve the Bogoliubov equations with the condensate wave function of Eq. \!(\ref{eq:cnds}) and calculate the excitation spectrum of the condensate in a Kronig-Penney potential.

\subsection{Single barrier problem}
A periodic potential can be viewed as a periodic array of potential barriers.
The band structure of a single particle in a periodic potential can be expressed in terms of the tunneling properties of the particle in the presence of a single barrier potential~\cite{rf:ashc}.
When one solves the Schr\"odinger equation with a single barrier, the wave function including the transmission amplitude is obtained.
By imposing the Bloch's theorem on the wave function, one obtains an equation to determine the band structure~\cite{rf:ashc}.

We shall determine the band structure of the excitation spectrum.
In a similar way to the case of a single particle, one can relate the band structure of the excitation spectrum to the tunneling properties of the excitations in the presence of a single potential barrier, as discussed in the next subsection.
In fact, once one solves the Bogoliubov equations with a single barrier and obtains the tunneling properties of the excitations, the band structure of the excitation spectrum can be easily obtained.
Hence, we first focus our attention on the $|x|<\frac{a}{2}$ region, and solve the Bogoliubov equations analytically, with regard to the single barrier problem.

Kagan {\it et al}. \!have solved the \!Bogoliubov equations with a single rectangular barrier and numerically calculated the transmission amplitude~\cite{rf:antun}.
Although they have also approximately obtained an analytical expression for the transmission amplitude, the approximation is not valid at very low energies.
Since behaviors of the transmission amplitude around zero energy are crucial to the form of the excitation spectrum in an optical lattice, we need to obtain another analytical expression for the transmission amplitude.
In the model of a $\delta$-function potential barrier which corresponds to the thin barrier limit of a rectangular barrier, one can calculate the transmission amplitude exactly.

\begin{figure}[t]
\includegraphics[width= 3.2 in, height= 1.5 in]{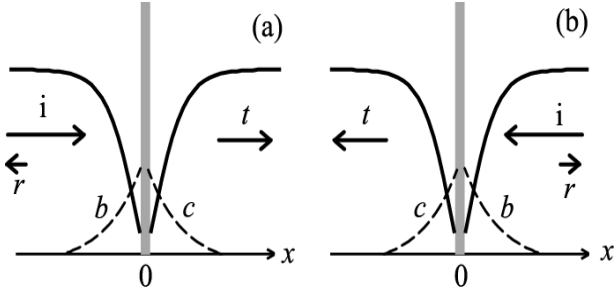}
\caption{\label{fig:sc_sol}
The schematic pictures of the solutions of the single barrier problem.
Scattering processes in which (a) the excitation comes from the left hand side, and (b) the excitation comes from the right hand side.
Solid curves, arrows and the dashed curves represent the condensate wave functions, the scattering components and the localized components, respectively.
                              }
\end{figure}

There exist two independent solutions of the single barrier problem, corresponding to the types of scattering process.
One solution ${\psi}^l(x)$ describes the process where a Bogoliubov excitation comes from left, and the other solution ${\psi}^r(x)$ describes the process where a Bogoliubov excitation comes from right.
The schematic pictures of the solutions of the single barrier problem are shown in Fig. \!\ref{fig:sc_sol}.
Substituting the condensate wave function of Eq. \!(\ref{eq:cnds}) into the Bogoliubov equations, let us obtain the solutions ${\psi}^l(x)$ and ${\psi}^r(x)$.
On both sides of the barrier, one obtains four particular solutions analytically~\cite{rf:antun} under the boundary condition at $|x|\gg\xi$:
  \begin{eqnarray}
    \left(u_n(x), \, v_n(x)\right)^{\bf t} \propto e^{\frac{ip_n x}{\hbar}}.
  \end{eqnarray}
These solutions are 
 \begin{eqnarray}
   u_n(x) &=& \Lambda_n e^{\frac{ip_n x}{\hbar}}\Biggl[\biggr.
                     \mathrm{tanh}\left(\frac{|x|+x_0}{\xi}\right)
                     -i\,\mathrm{sgn}(x) \frac{p_n\xi}{2\hbar\varepsilon} 
                     \nonumber \\
          & & \times\left(\varepsilon+\mu-\mu\,
              \mathrm{tanh}^2\left(\frac{|x|+x_0}{\xi}\right)\right)
                     \label{eq:ansltu} \\
           & & +\frac{E_{p_n}}{\varepsilon}
                     \mathrm{tanh}\left(\frac{|x|+x_0}{\xi}\right)
                     -i\,\mathrm{sgn}(x)\frac{E_{p_n}p_n\xi}{2\hbar\varepsilon}
                     \biggl.\Biggr],\nonumber
                      \\
   v_n(x) &=& \Lambda_n e^{\frac{ip_n x}{\hbar}}\Biggl[\Biggr.
                     \mathrm{tanh}\left(\frac{|x|+x_0}{\xi}\right)
                     -i\,\mathrm{sgn}(x) \frac{p_n\xi}{2\hbar\varepsilon} 
                     \nonumber \\
          & & \times\left(\varepsilon-\mu+\mu\,
              \mathrm{tanh}^2\left(\frac{|x|+x_0}{\xi}\right)\right)
                     \label{eq:ansltv} \\
          & & -\frac{E_{p_n}}{\varepsilon}
              \mathrm{tanh}\left(\frac{|x|+x_0}{\xi}\right)
                     +i\,\mathrm{sgn}(x)\frac{E_{p_n}p_n\xi}{2\hbar\varepsilon}
                     \Biggl.\Biggr],
                     \nonumber               
   \end{eqnarray}
where
  \begin{eqnarray} 
     p_{1,2} &=& \pm p = \pm \sqrt{2m(\sqrt{\mu^2+\varepsilon^2}-\mu)}, 
     \label{eq:scat}\\
     p_{3,4} &=& \mp i\gamma=\mp i\sqrt{2m(\sqrt{\mu^2+\varepsilon^2}+\mu)},\, 
     \label{eq:locl}
  \end{eqnarray}
  \begin{eqnarray}
     \Lambda_n &=&  \sqrt{\frac{\mu^2}{2g\varepsilon}} \times 
       \left\{\!\begin{array}{cc}
         \frac{\sqrt{2\mu}+\mathrm{sgn}(x)\,i\sqrt{E_p}}{\sqrt{2\mu+E_p}},
         \,\,\, 
         n=1, \\
         \frac{\sqrt{2\mu}-\mathrm{sgn}(x)\,i\sqrt{E_p}}{\sqrt{2\mu+E_p}},
         \,\,\, 
         n=2, \\
         1, \,\,\,\,\,\,\,\,\,n=3, 4
                  \end{array}\!\right.,
         \\
         E_{p_n} &=& \frac{p_n^2}{2m}.
  \end{eqnarray}
Wave functions $\left(u_1(x), v_1(x)\right)^{\bf t}$ and $\left(u_2(x), v_2(x)\right)^{\bf t}$ describe scattering components.
Wave functions $\left(u_3(x), v_3(x)\right)^{\bf t}$ at $x<0$ and $\left(u_4(x), v_4(x)\right)^{\bf t}$ at $x>0$ describe the localized components around the potential barrier, because they decay exponentially at $|x|\gg\xi$.
Wave functions $\left(u_3(x), v_3(x)\right)^{\bf t}$ at $x>0$ and $\left(u_4(x), v_4(x)\right)^{\bf t}$ at $x<0$ diverge far from the potential barrier.
It is noted that Eqs. \!(\ref{eq:scat}) and (\ref{eq:locl}) can be derived by solving
  \begin{eqnarray}
    \varepsilon=\sqrt{\frac{p_n^2}{2m}(\frac{p_n^2}{2m}+2\mu)}.
    \label{eq:hg_bg}
  \end{eqnarray}
Equation \!(\ref{eq:hg_bg}) expresses the Bogoliubov spectrum for a uniform system.
The normalization constant $\Lambda_n$ of the scattering components is determined to satisfy
  \begin{equation}
    u_n(x), \, v_n(x) = \sqrt{\frac{E_p+\mu\pm\varepsilon}{2\varepsilon}}
                          e^{\frac{ip_n x}{\hbar}} 
  \end{equation}
at $|x|\gg\xi$.

We omit the unphysical divergent components.
The solutions ${\psi}^l(x)$ and ${\psi}^r(x)$ are expressed as superpositions of the remaining components.
The solution ${\psi}^l(x)$ is written as
\begin{eqnarray}
     {\psi}^l(x)=
     \left(\!\begin{array}{cc}
                      u^l \\ v^l
                  \end{array}\!\right)
\!\!\!&=&\!\!\!\left\{\begin{array}{ll}
          \left(\!\begin{array}{cc}
                    u_1 \\ v_1
                         \end{array}\!\right)
          +r\left(\!\begin{array}{cc}
                    u_2 \\ v_2
                         \end{array}\!\right)\\ 
          +b\left(\!\begin{array}{cc}
                    u_3 \\ v_3
                         \end{array}\!\right),
                         & \!\!\!x<0,  \\
          t\left(\!\begin{array}{cc}
                    u_1 \\ v_1
                         \end{array}\!\right)
          +c\left(\!\begin{array}{cc}
                    u_4 \\ v_4
                         \end{array}\!\right),
                         & \!\!\!x>0,
          \end{array}\right.\label{eq:lcs}
\end{eqnarray}
where the coefficients $r$, $b$, $t$, and $c$ are the amplitudes of the reflected, the left localized, the transmitted, and the right localized components, respectively. 
They are functions of the energy $\varepsilon$ and the potential strength $V_0$.
The boundary conditions at $x=0$ yield four equations to determine all the coefficients
  \begin{eqnarray}
     {\psi}^l(+0)={\psi}^l(-0),\label{eq:bc}
  \end{eqnarray}
  \begin{eqnarray}
     \left. \frac{d{\psi}^l}{dx} \right|_{+0}\!\!\!
     =\left. \frac{d{\psi}^l}{dx} \right|_{-0}\!\!\!
             +\frac{2mV_0}{\hbar^2}{\psi}^l(0). 
     \label{eq:bcde} 
  \end{eqnarray}
These equations are linear simultaneous equations for the coefficients $r$, $b$, $t$ and $c$, and one can analytically solve them.
Since the exact solutions of Eqs. \!(\ref{eq:bc}) and (\ref{eq:bcde}) are unnecessarily complicated, we only write approximate forms of the coefficients.
When $\varepsilon \ll \mu$ and $V_0 \gg \mu\xi$, one approximately obtains all the coefficients as
  \begin{eqnarray}
       r &=& \frac{\varepsilon V_0-\varepsilon\mu\xi
             +i\frac{\varepsilon^2 V_0}{\mu}}{Z},
               \\
       b &=& -c = \frac{-2\varepsilon\mu\xi+\frac{4\varepsilon\mu^2\xi^2}{V_0}
                +i\varepsilon^2\xi}{4Z},
               \\
       t &=& \frac{\varepsilon\mu\xi+i\mu^2\xi}{Z}, \,
       \label{eq:cfct}\\
       Z &=& \varepsilon V_0-\varepsilon\mu\xi+i\mu^2\xi.\label{eq:denom}
  \end{eqnarray}
Thus, we obtained $\psi^l(x)$ analytically, and we can also calculate $\psi^r(x)$ in the same way.
\begin{figure}[t]
\includegraphics[width=2.8 in, height=1.8 in]{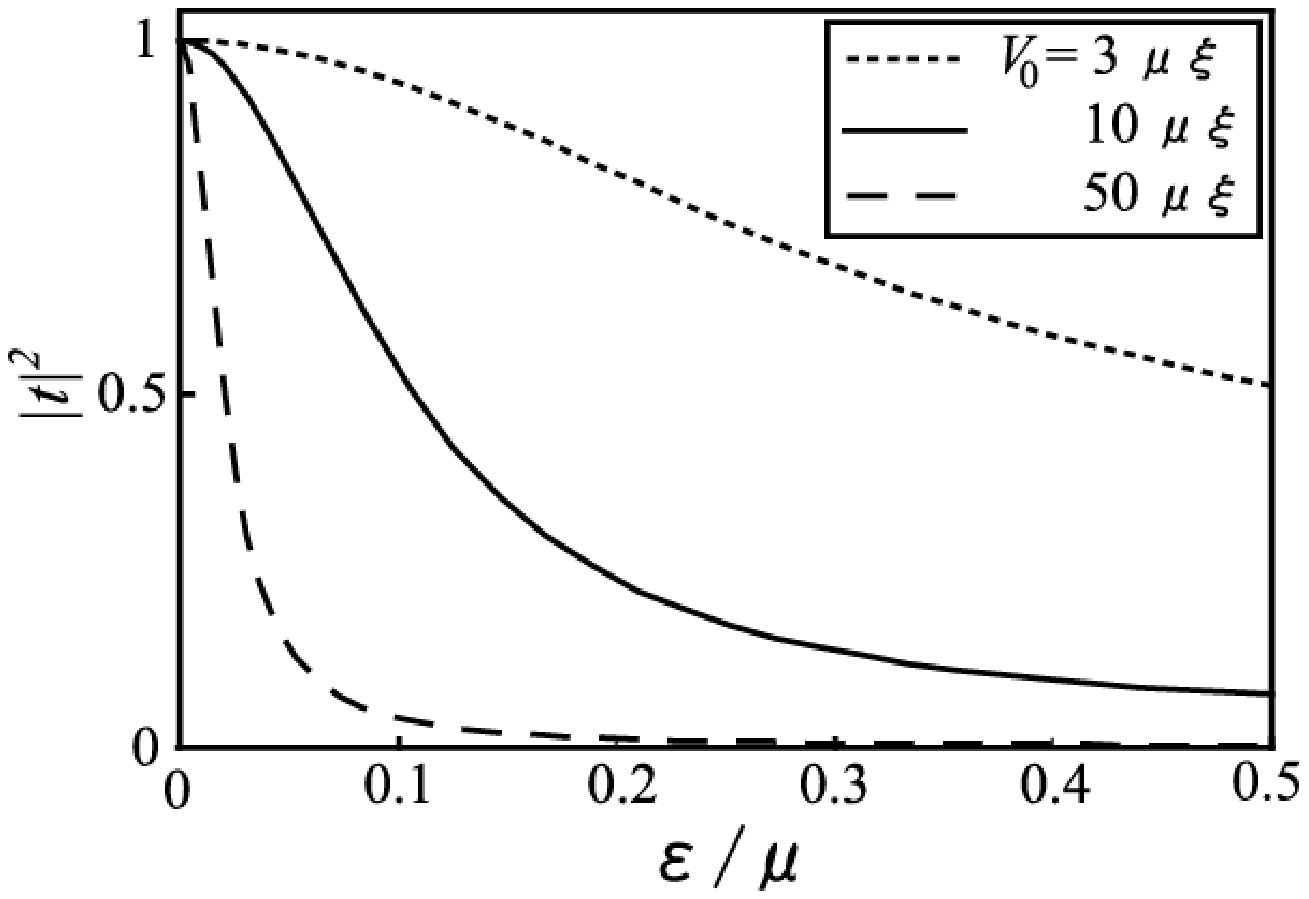}
\caption{\label{fig:prob_tun}
The transmission coefficient $|t|^2$ with $V_0=3 \mu\xi$ (dotted lines), $V_0=10 \mu\xi$ (solid lines) and $V_0=50 \mu\xi$ (dashed lines), as functions of the energy.}
\end{figure}
\begin{figure}[t]
\includegraphics[width=2.8in, height=1.8in]{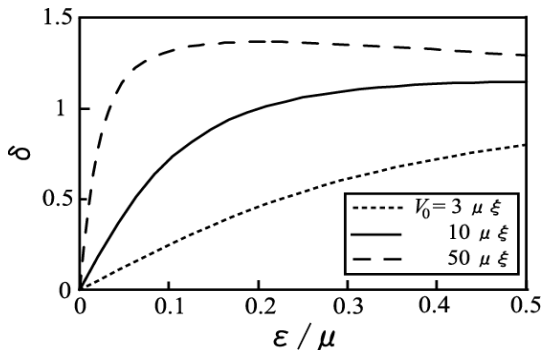}
\caption{\label{fig:phase_tun}
The phase shift $\delta$ of $t$ with $V_0=3 \mu\xi$ (dotted lines), $V_0=10 \mu\xi$ (solid lines) and $V_0=50 \mu\xi$ (dashed lines), as functions of the energy.}
\end{figure}

The transmission coefficient $T \equiv |t|^2$ and the phase shift $\delta\equiv {\rm arg}(t)$ are shown in Figs. \ref{fig:prob_tun} and \ref{fig:phase_tun}, respectively, as a function of the energy.
Expanding $t$ around $\varepsilon=0$, one can analytically obtain approximate expressions of $|t|$ and $\delta$
  \begin{eqnarray}
    |t| &\simeq& 1-\alpha\left(\frac{\varepsilon}{\mu}\right)^2,
    \label{eq:unexp_pr}\\ 
    \delta &\simeq& \beta\frac{\varepsilon}{\mu}.
    \label{eq:tnexp_ph}
  \end{eqnarray}
The coefficients $\alpha$ and $\beta$ are
  \begin{eqnarray}
  &&\!\!\!\!\!\!\!\!\!\!\!\!\!
  \alpha=
  \nonumber\\
  &&\!\!\!\!\!\!\!\!\!\!\!\!\!
  \frac{2(V_0 \!-\! \mu\xi)(V_0^3\!+\! \nu V_0^2\!+\! 2\nu(\mu\xi)^2
  \!-\! 4(\mu\xi)^3)\!+\! 9(\mu\xi V_0)^2}{8(\mu\xi\nu)^2}\!,
  \end{eqnarray}
  \begin{eqnarray}
  \beta=\frac{V_0^2+\nu V_0-3\mu\xi\nu+6(\mu\xi)^2}{2\mu\xi\nu},
  \end{eqnarray}
where
  \begin{eqnarray}
  \nu=\sqrt{V_0^2+4(\mu\xi)^2}.
  \end{eqnarray}
It is obvious from Figs. \!\ref{fig:prob_tun} and \ref{fig:phase_tun} and Eqs. \!(\ref{eq:unexp_pr}) and (\ref{eq:tnexp_ph}) that the transmission coefficient $T$ approaches unity and the phase shift $\delta$ approaches zero as the energy is reduced to zero.
This means that the potential barrier is transparent for low energy excitations.
This behavior of the low energy excitations has been called the {\it anomalous tunneling} by Kagan $\mathit{et \,\, al.}$~\cite{rf:antun}.
Equations \!(\ref{eq:cfct}) and (\ref{eq:denom}) shows that the peak of $T$ has a Lorentzian shape with half width $\Delta\varepsilon\sim V_0^{-1}$, and the anomalous tunneling becomes restricted to only the excitations with very low energies as strength of the potential barrier increases.
This peculiar tunneling behavior of the excitations appears for arbitrary values of the potential strength.

\subsection{Band structure of excitation spectrum}
In this subsection, we analytically calculate the band structure of the excitation spectrum with use of the solution of the single barrier problem obtained in the previous subsection.
We show that the tunneling properties of the excitations determine the band structure.

Since the Bogoliubov equations are linear differential equations, a general solution of the equations can be described as a linear combination of independent solutions with the same energy.
We can write a general solution of the Bogoliubov equations in the region $|x|<\frac{a}{2}$ as a linear combination of $\psi^l(x)$ and $\psi^r(x)$~\cite{rf:footnote}:
  \begin{eqnarray}
    \psi(x) = \chi \psi^l(x) + \zeta \psi^r(x), \,\,\,\,\,|x|<\frac{a}{2}.
    \label{eq:gene}
  \end{eqnarray}
where $\chi$ and $\zeta$ are arbitrary constants.

Now we extend this solution to all regions of $x$ by means of the Bloch's theorem.
The Bloch's theorem asserts that $\psi$ satisfies
  \begin{eqnarray}
    \psi(x+a) &=& e^{\frac{iqa}{\hbar}} \psi(x),\label{eq:bloch}\\
    \left.\frac{d\psi}{dx}\right|_{x+a} &=&
    e^{\frac{iqa}{\hbar}} \left.\frac{d\psi}{dx}\right|_{x},
    \label{eq:bloch_de}
  \end{eqnarray}
where $q$ is the quasi-momentum.
Equations \!(\ref{eq:bloch}) and (\ref{eq:bloch_de}) at $x=-\frac{a}{2}$ yields an equation expressing the relation between the excitation energy $\varepsilon$ and the quasi-momentum $q$:
  \begin{eqnarray}
    \mathrm{cos}\left(\frac{qa}{\hbar}\right) 
    = \frac{t^2-r^2}{2\,t}e^{\frac{ipa}{\hbar}}
      +\frac{1}{2\,t}e^{-\frac{ipa}{\hbar}}.
    \label{eq:relek}
  \end{eqnarray}
Since we are assuming $a\gg\xi$, the localized components around the potential barriers, which decay exponentially and vanish at $|x|=\frac{a}{2}$, do not appear explicitly in Eq. \!(\ref{eq:relek}).

The Wronskian defined as
  \begin{eqnarray}
    \!W\!(\psi^{j\ast}\!, \!\psi^i)\!=\!
    u^{j\ast}\frac{d}{dx}u^{i}\!-\!u^i\frac{d}{dx}u^{j\ast}\!
    +\!v^{j\ast}\frac{d}{dx}v^i\!-\!v^i\frac{d}{dx}v^{j\ast}
  \end{eqnarray}
yields relations between $r$ and $t$.
We can simplify Eq. \!(\ref{eq:relek}) by the relations.
One can easily prove from the Bogoliubov equations that $W$ is independent of $x$ when $\psi^j$ and $\psi^i$ have the same energy.
By evaluating $W(\psi^{l\ast}, \psi^l)$, one obtains the conservation law of the energy flux~\cite{rf:antun}:
  \begin{eqnarray}
    |t|^2+|r|^2=1.\label{eq:efcl}
  \end{eqnarray}
Meanwhile, by evaluating $W(\psi^{r\ast}, \psi^l)$, one obtains the relation
  \begin{eqnarray}
    t=|t|e^{i\delta}, r=\pm i|r|e^{i\delta}.\label{eq:prim}
  \end{eqnarray}
Substituting Eqs. \!(\ref{eq:efcl}) and (\ref{eq:prim}) into Eq. \!(\ref{eq:relek}), one obtains the simplified relation between the excitation energy and the quasi-momentum
  \begin{eqnarray}
    \frac{\mathrm{cos}\left(\frac{pa}{\hbar}+\delta\right)}{|t|}=
    \mathrm{cos}\left(\frac{qa}{\hbar}\right) \, .\label{eq:sirel}
  \end{eqnarray}
This relation is exactly the same form as that in the case of a single particle~\cite{rf:ashc}.
We clearly see from Eq. \!(\ref{eq:sirel}) that the tunneling properties in the single barrier problem determine the relation between the excitation energy and the quasi-momentum, namely the band structure.
\begin{figure}[t]
\includegraphics[width=3in, height=2.2in]{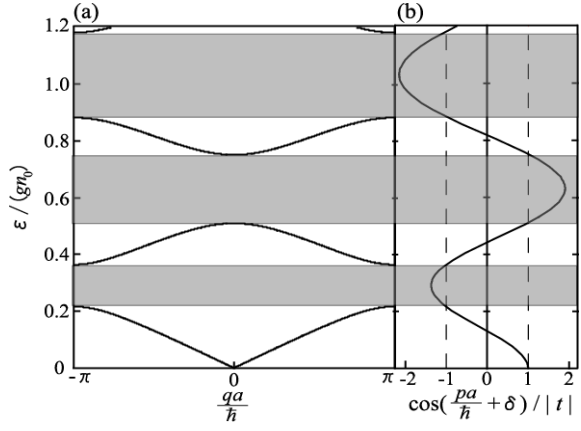}
\caption{\label{fig:epsart}
The band structure of the excitation spectrum in the presence of a Kronig-Penney potential with $a=10\xi_0$ and $V_0=5 gn_0\xi_0$ are shown.
Shaded areas express band gaps due to the absence of solutions of Eq. \!(\ref{eq:sirel}).
(a) The band structure is shown.
(b) The left-hand side of Eq. \!(\ref{eq:sirel}) as a function of energy is shown. 
                 }
   \end{figure}

Solving Eq. \!(\ref{eq:sirel}) for the excitation energy $\varepsilon$, we obtain the band structure of the excitation spectrum as shown in Fig. \!\ref{fig:epsart}(a).
Fig. \!\ref{fig:epsart}(b) shows the left-hand side of Eq. \!(\ref{eq:sirel}) as a function of $\varepsilon$.
There exists no solution of Eq. \!(\ref{eq:sirel}) when the absolute value of the left-hand side exceeds unity, because the absolute value of the right-hand side is equal to or less than unity.
Therefore, the energy regions where the absolute value of the left-hand side is greater than unity correspond to the forbidden regions of the excitation energy, namely the band gaps.
They are expressed as the shaded regions in Fig. \!\ref{fig:epsart}.

In the strong potential limit $V_0\to\infty$, the widths of all the energy bands become narrow.
The energy bands approach the discrete eigenenergies of the Bogoliubov equations for a single potential well, because the condensate is perfectly divided into each well.
\subsection{Phonon dispersion}
Fig. \!\ref{fig:epsart}(a) shows that the first band of the excitation spectrum is phonon-like in the low energy regions.
We shall verify that the form of the excitation spectrum at low energies is phonon-like for arbitrary values of the potential depth.
The excitation spectrum in the low energy limit $\varepsilon\to0$ is calculated by expanding Eq. \!(\ref{eq:sirel}) around $\varepsilon=0$.
Substituting $p \simeq \sqrt{\frac{m}{\mu}}\varepsilon$, Eqs. \!(\ref{eq:unexp_pr}) and (\ref{eq:tnexp_ph}) into Eq. \!(\ref{eq:sirel}), one obtains the phonon dispersion of the excitation spectrum
  \begin{eqnarray}
  \varepsilon \simeq cq,
  \end{eqnarray}
where the phonon velocity $c$ is 
  \begin{eqnarray}
  c=\sqrt{\frac{\mu a^2}{m\left((a+\beta\xi)^2-2\xi^2\alpha\right)}}.
  \label{eq:phonov}
  \end{eqnarray}
When $V_0\gg gn_0\xi_0$, Eq. \!(\ref{eq:phonov}) can be approximated as
  \begin{eqnarray}
    c \simeq
    c_0\sqrt{\frac{gn_0a}{2V_0+gn_0a}}
    \left(1+\frac{5\xi_0}{2a}\right),\label{eq:phonovl}
  \end{eqnarray}
where $c_0\equiv\sqrt{\frac{gn_0}{m}}$ is the phonon velocity in uniform systems.
When $V_0\ll gn_0\xi_0$, Eq. \!(\ref{eq:phonov}) can be approximated as
  \begin{eqnarray}
    c \simeq
    c_0\left(1-\frac{3V_0^2}{16a\xi_0(gn_0)^2}\right).
  \end{eqnarray}

\begin{figure}[t]
      \includegraphics[width=2.8in, height=1.8in]{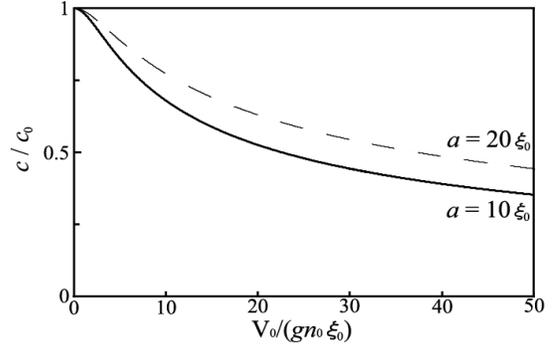}
         \caption{\label{fig:epsart2}
            The phonon velocities of Eq. \!(\ref{eq:phonov})
            in Kronig-Penney potentials with $a=10 \xi_0$ and $20 \xi_0$,
            as a function of the potential strength $V_0$.
                              }
   \end{figure}

The phonon velocities for $a=10 \xi_0$ and $20 \xi_0$ as functions of $V_0$ are shown in Fig. \!\ref{fig:epsart2}.
The phonon velocity decreases monotonically as the barrier strength increases, and this behavior is qualitatively consistent with the case of a sinusoidal potential~\cite{rf:comp}.

Thus the anomalous tunneling properties of Eqs. \!(\ref{eq:unexp_pr}) and (\ref{eq:tnexp_ph}), namely the perfect transmission of very low energy excitations, leads to the phonon dispersion.
It is pointed out in Ref.~\cite{rf:soft} that the excitation spectrum shows softening at large wavelengths for a deep sinusoidal lattice potential, relating to the instability of the superfluidity.
In contrast, the excitation spectrum is always phonon-like in the Kronig-Penney potential because the anomalous tunneling occurs even for very strong potential barriers.
The phonon dispersion of the excitation spectrum at low energies reflects the superfluidity of the condensate.
In this sense, the anomalous tunneling is crucial to the stability of the superfluidity in the Kronig-Penney potential.

However, our results cannot eliminate the possibility of the softening in a sinusoidal periodic potential, because we have assumed in our calculation that the lattice constant is sufficiently larger than the healing length.
In the Kronig-Penney model, this assumption assures that the size of the condensate wave function in each well hardly changes as the lattice potential becomes deep; consequently the size of the condensate is always sufficiently larger than the healing length.
In contrast, even when the lattice constant is sufficiently larger than the healing length, the size of the condensate wave function in each well can be comparable to or smaller than the healing length in a deep sinusoidal potential.
This is because the deeper the sinusoidal lattice potential is, the tighter the confinement of each well is.
Thus, a possibility of the softening is remaining in the situation where the size of the condensate in each well is comparable to or smaller than the healing length.
   \begin{figure}[t]
      \includegraphics[width=2.8in, height=2in]{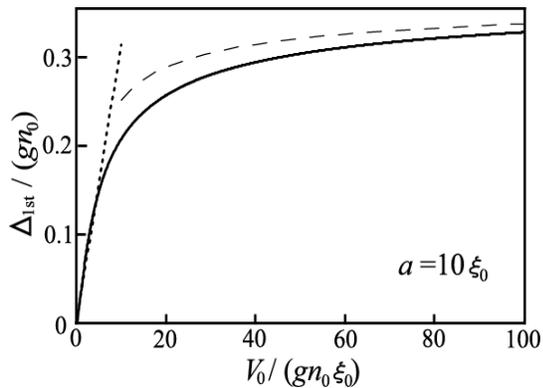}
         \caption{\label{fig:epsart3}
            The energy band gap between the lowest and the second lowest band
            for $a=10\xi_0$,
            as a function of the potential strength $V_0$.
            The dashed (dotted) line shows an approximate value of the band gap
            in the limit of $V_0\gg gn_0 a$ ($V_0\ll gn_0\xi_0$).
                  }
   \end{figure}
\subsection{Band gap}
The $j$th band gap $\Delta_{j\mathrm{th}}$ is determined by the excitation energies of the top of lower band and the bottom of higher band, and they are obtained by solving Eq. \!(\ref{eq:sirel}) at $q=\pm\frac{\hbar\pi}{a}$.
Here we show the first band gap $\Delta_{1{\rm st}}$ between the lowest and the second lowest band, as a function of the potential strength $V_0$ in Fig. \!\ref{fig:epsart3}.

When $V_0\gg gn_0 a$, the first band gap is much larger than the first band width.
In this case, one can approximately obtain
  \begin{eqnarray}
    \Delta_{\mathrm{1st}}\!\simeq\!
    \frac{\pi gn_0\xi_0}{a}\!\left(1+\frac{2\xi_0}{a}\right)
    \!-\!gn_0\sqrt{\frac{gn_0\xi_0^2}{V_0a}}
    \!\left(1+\frac{5\xi_0}{2a}\right)\! .
    \label{eq:gapl}    
  \end{eqnarray}
When $V_0 \ll gn_0\xi_0$, one can obtain
  \begin{eqnarray}
    \Delta_{\mathrm{1st}}\simeq
        \frac{\pi \xi_0 V_0}{a^2}.\label{eq:gaps}
  \end{eqnarray}
Equation \!(\ref{eq:gaps}) and Fig. \!\ref{fig:epsart3} show that at first the band gap increases linearly with the potential strength.
The first term of the right-hand side of Eq. \!(\ref{eq:gapl}) corresponds to the lowest excitation energy of a condensate in a single well.
Since the second term vanishes in the limit of $V_0\to\infty$, the band gap becomes the lowest excitation energy in a single well.
\subsection{Tight-binding limit}
When the potential is sufficiently strong $V_0 \gg gn_0 a$, one can approximately solve Eq. \!(\ref{eq:sirel}) and obtain the first band of the excitation spectrum.
Since the excitation energy in the first band is much smaller than the chemical potential in this situation, the approximate expression of $t$ of Eqs. \!(\ref{eq:cfct}) and (\ref{eq:denom}) is adequate.
One can rewrite Eqs. \!(\ref{eq:cfct}) and (\ref{eq:denom}) as
  \begin{eqnarray}
   {\rm cos}\,\delta &=& |t|\left(1+\frac{\varepsilon^2 V_0}{\mu^3 \xi}\right),
   \label{eq:phs_cos}\\
   {\rm sin}\,\delta &=& 
   |t|\frac{\varepsilon}{\mu}\left(\frac{V_0}{\mu\xi}-2\right),
   \label{eq:phs_sin}
  \end{eqnarray}
where
  \begin{eqnarray}
   |t|=\frac{\mu^2 \xi}{\sqrt{\varepsilon^2 V_0(V_0-2\mu\xi)+\mu^4 \xi^2}}.
   \label{eq:abs_tun}
  \end{eqnarray}
Substituting Eq. \!(\ref{eq:phs_cos}), (\ref{eq:phs_sin}) and (\ref{eq:abs_tun}) into Eq. \!(\ref{eq:sirel}), one obtains
  \begin{equation}
   \varepsilon\simeq
   gn_0\sqrt{\frac{2gn_0\xi_0^2}{aV_0}}\left(1+\frac{5\xi_0}{2a}\right)
   \left|\mathrm{sin}\left(\frac{qa}{2\hbar}\right)\right|,\label{eq:loweb}
  \end{equation}
where $\mu$ is replaced with $gn_0$ by using Eq. \!(\ref{eq:cheml}).

Meanwhile, the first band of the excitation spectrum has been calculated using the tight-binding approximation in many theoretical papers~\cite{rf:tight,rf:DNLS,rf:comp,rf:finite}.
For a very deep lattice, the spectrum takes the form
  \begin{equation}
    \varepsilon \simeq \sqrt{\frac{2\eta}{\kappa}}
    \left|\mathrm{sin}\left(\frac{qa}{2\hbar}\right)\right|,\label{eq:ttba}
  \end{equation}
where $\kappa$ is the compressibility of the system and $\eta$ is the lowest energy bandwidth of stationary current-carrying states of the condensate~\cite{rf:comp}.
The compressibility $\kappa$ is defined by the thermodynamic relation
  \begin{equation}
    \frac{1}{\kappa}=N_0\frac{\partial\mu}{\partial N_0}.\label{eq:cmprs}
  \end{equation}
The parameter $\eta$ is related to the effective mass $m^{\ast}$ through the relation
  \begin{equation}
    \eta=\frac{2\hbar^2}{m^{\ast}a^2}.\label{eq:bwdth}
  \end{equation}
The definition of the effective mass is
  \begin{equation}
    \frac{1}{m^{\ast}}=\frac{\partial^2 \tilde{\epsilon}}{\partial k^2}
                       \Bigl.\biggr|_{k=0}
  \end{equation}
where $k$ is the quasi-momentum of the current-carrying condensate and $\tilde{\epsilon}$ is the energy per particle of the condensate.
We should be careful not to confuse the energy dispersion of the Bogoliubov excitations $\varepsilon(q)$, which is our major interest, and the energy per particle $\tilde{\epsilon}(k)$ of the condensate itself.

Using Eqs. (\ref{eq:cmprs}) and (\ref{eq:bwdth}), one can obtain
  \begin{eqnarray}
    \frac{1}{\kappa}&\simeq&
    gn_0\left(1+\frac{\xi_0}{a}\right),\label{eq:concc}\\
    \eta&\simeq&\frac{(gn_0\xi_0)^2}{aV_0}\left(1+\frac{4\xi_0}{a}\right).
    \label{eq:concb}
  \end{eqnarray}
Here we have used Eq. \!(\ref{eq:cheml}) to calculate the compressibility.
Substituting Eqs. \!(\ref{eq:concc}) and (\ref{eq:concb}) into Eq. \!(\ref{eq:ttba}), we see that our calculation of the first band of the excitation spectrum is consistent with the result of the tight-binding approximation when the potential is sufficiently strong.

\section{CONCLUSION}
In summary, we have investigated the band structure of the excitation spectrum of the condensate in a Kronig-Penney potential.
Connecting the analytical solutions of Bogoliubov equations in the single barrier problem by means of the Bloch's theorem, the relation between the excitation energy and the quasi-momentum have been obtained, which determines the band structure.
We have shown that the excitation spectrum is gapless and linear in the low energy region due to the anomalous tunneling property of the excitations.

While we have considered current-free condensates in our calculations, the elementary excitations of condensates with large current in periodic potentials are known to exhibit the Landau and dynamical instabilities~\cite{rf:dynam,rf:DNLS,rf:inst,rf:dy_inst}.
Since both instabilities are associated with the specific properties of the excitation spectra in the low energy region, the tunneling properties at low energy may affect the instability when the problem is considered with the Kronig-Penney model.
Hence, it will be interesting to study the excitations of the current-carrying condensates in the Kronig-Penney potential and discuss the relation between the instability and the anomalous tunneling of the Bogoliubov excitations.

{\it Note added.} After the submission of this paper, we learned of a relevant paper by Kovrizhin~\cite{rf:add}.
In Ref.~\cite{rf:add}, the tunneling problem of the Bogoliubov excitations through a $delta$-function potential barrier is investigated.
Exact form of the wave function Eq. \!(\ref{eq:lcs}) is derived, and the anomalous behavior of the transmission coefficient is discussed.
\begin{acknowledgments}
The authors would like to thank T. Kimura, N. Yokoshi, T. Nikuni and N. Hatano for fruitful comments and discussions.
The work is partially supported by a Grant for the 21st Century COE Program (Holistic Research and Education center for Physics  of Self-organization 
Systems) at Waseda University from the Ministry of Education, Sports, Culture, Science and Technology of Japan.
I. D. is  supported by Grant-in-Aid for JSPS fellows.

\end{acknowledgments}


\begin{thebibliography}{99}
\bibitem{rf:footnote}
Since one can consider the Bogoliubov equations as a linear differential equation for $u(x)$ of fourth order, there should be four independent solutions with the same eigenenergy.
Two solutions of them are $\psi^l(x)$ and $\psi^r(x)$.
The other two solutions are the unphysical divergent components omitted in subsection III. A, and they are also omitted in Eq. \!(\ref{eq:gene}).
Consequently, the solution of Eq. \!(\ref{eq:gene}) is a superposition of $\psi^l(x)$ and $\psi^r(x)$.
\end{thebibliography}
\end{document}